\documentclass[aps,showpacs,preprintnumbers,amsmath,prd,amssymb,superscriptaddress]{revtex4}
\usepackage{amssymb}
\usepackage{CJK}
\usepackage{amsmath,amssymb,graphicx,bm}

\begin{document}

\title{Large-scale structure in superfluid Chaplygin gas cosmology}

\author{Rongjia Yang\footnote{Corresponding author}}
\email{yangrongjia@tsinghua.org.cn}
\affiliation{College of Physical Science and Technology, Hebei University, Baoding 071002, China}
\affiliation{Kavli Institute for Theoretical Physics China, Institute of Theoretical Physics, Chinese Academy of Science, Beijing 100190, China}
\affiliation{Center for High Energy Physics, Peking University, Beijing 100871, China}

\begin{abstract}
We investigate the growth of large-scale structure in the superfluid Chaplygin gas (SCG) model. Both linear and nonlinear growth, such as $\sigma_8$ and the skewness $S_3$, are discussed. We find the growth factor of SCG reduces to the Einstein¨Cde Sitter case at early times while it differs from the cosmological constant model ($\Lambda$CDM) case in the large $a$ limit. We also find there will be more stricture growth on large scales in the SCG scenario than in $\Lambda$CDM and the variations of $\sigma_{8}$ and $S_3$ between SCG and $\Lambda$CDM cannot be discriminated.
\end{abstract}

\makeatletter
\def\@pacs@name{PACS: 95.36.+x, 98.65.Dx, 98.80.-k}
\makeatother

\maketitle

\section{Introduction}
A wealth of astronomical observations indicated that the Universe is experiencing accelerated expansion. Assuming the validity of general relativity on large scales, an unknown energy component, the dark energy, is usually introduced to explain the accelerated expansion.
The simplest candidate is the cosmological constant model ($\Lambda$CDM) that is consistent
with most of the current astronomical observations but suffers from the cosmological constant problem \cite{weinberg} and age problem \cite{Yang2010}.
It is thus natural to study more complicated cases, such as models allowing a time evolution of the dark energy component: quintessence, phantom, k-essence, quintom, tachyon, and other scalar fields. Recently, superfluid Chaplygin gas (SCG), a model unified dark matter and dark energy, was proposed \cite{Popov}. SCG involves the Bose-Einstein condensate as dark energy possessing the equation of state (EoS) of Chaplygin gas, and an excited state acts as dark matter. Though the condensate behaves like Chaplygin gas \cite{Kamenshchik}, the evolution of the Universe provided by SCG is different from that in the model with Chaplygin gas as dark energy as well as from that in the model with Chaplygin gas unified dark matter and dark energy \cite{Popov}. In Chaplygin gas unified dark matter and dark energy, the negligible sound speed that produces unphysical oscillations and an exponential blowup in the dark matter power spectrum at present \cite{Sandvik}. This problem was solved for the generalized Chaplygin gas model by decomposing the energy density into dark matter and dark energy \cite{Bento}. Recently, observational constraints on the generalized Chaplygin model have been considered in \cite{Campos}. SCG has been analyzed from the point of view of statefinder \cite{Popov2011} and has been constrained by using observational data \cite{Lazkoz}.

Dark energy can affect the expansion rate causing geometrical effects that can be revealed though distance measurements, and can also affect structure formation that can be quantified by the growth factor. Thus, structure formation will be affected by the amount of dark energy and by its dynamical evolution over cosmological history. In this paper we study the large-scale structure growth in the SCG scenario from the point of view of gravitational collapse. The linear and nonlinear growth can be related to observations and hence lead to observational constraints, such as the skewness of the density field and the rms fluctuations on a sphere of 8 Mpc $h^{-1}$ \cite{Dev}. We probe the significance of the SCG dark energy component throughout the epoch during which large-scale structures grow.

The rest of the paper is organized as follows. In Sec. II, we sketch the derivation of the spherical collapse formalism. In Sec. III, we review the SCG scenario. In Sec. IV, we investigate gravitational growth and the large structure in SCG cosmology. Conclusions and discussions are presented in Sec. IV.

\section{Gravitational Collapse}
To study the gravitational growth of the large-scale structure in SCG, we first briefly describe the usual formalism. The nonlinear equation for dust has been used in the context of structure formation \cite{Padmanabhan, Abramo} and for the study of the spherical and ellipsoidal collapse \cite{Bernardeau, Ohta}. The linearized equation was presented in \cite{Coles} for dust and relativistic matter and in \cite{Lima} for a general model with constant EoS. The spherical collapse model has been considered in \cite{Pace,Wang} for several dark-energy scenarios and in \cite{Multamaki2003,Multamaki} for nonstandard cosmologies and generalized Chaplygin gas (GCG) dark energy.

We consider a homogeneous and isotropic Friedmann-Lema\^{i}tre-Robertson-Walker (FLRW) universe with scalar factor $a$,
\begin{align}
ds^2=-dt^2+a^2(t)\left[\frac{dr^2}{1+Kr^2}+r^2(d\theta^2+\sin^2\theta d\phi^2)\right],
\end{align}
where the spatial curvature constant $K=+1$, 0, and $-1$ correspond to a closed, flat, and open universe, respectively. We define $k^2=8 \pi G$ and use the units
$c=1$ throughout this paper.

To obtain the equation of motion for the density contrast, $\delta=\rho/\overline{\rho}-1$ with $\overline{\rho}$ the mean background energy density, we resort to the Raychaudhuri equation which in general takes the form
\begin{align}
\dot{\Theta}+\frac{1}{3}\Theta=\omega_{\mu\nu}\omega^{\mu\nu}-\sigma_{\mu\nu}\sigma^{\mu\nu}+R_{\mu\nu}u^{\mu}u^{\nu},
\end{align}
where $u^\mu$ is the fluid's 4-velocity, $\Theta\equiv \nabla^{\mu}u_{\mu}$, $\sigma_{\mu\nu}$ is the shear tensor, and $\omega_{\mu\nu}$ is the vorticity tensor. For a shear-free and nonrotating fluid, $\sigma_{\mu\nu}=0$ and $\omega_{\mu\nu}=0$, which is the case we are interested in here, and assuming that the geometry of the Universe is of the FLRW form and then $\Theta=3H$ with $H$ the local Hubble rate, we get
\begin{align}
\label{Rmn1}
\dot{\Theta}+\frac{1}{3}\Theta^2=3(H^2+\dot{H}).
\end{align}

Choosing the coordinate system such that the 4-velocity of the fluid is $u^{\mu}=(1, \dot{a}\vec{x}+\vec{\upsilon})$ with $\vec{\upsilon}$ the peculiar velocity, we have
\begin{align}
\Theta\equiv \nabla^{\mu}u_{\mu}=\frac{1}{a}\nabla\cdot(\dot{a}\vec{x}+\vec{\upsilon})=3\frac{\dot{a}}{a}+\frac{\theta}{a}=3\overline{H}+\frac{\theta}{a},
\end{align}
where $\theta\equiv\nabla\cdot\vec{\upsilon}$ and $\overline{H}$ is the background Hubble rate. $\nabla_x=\frac{1}{a}\nabla_X$ with $X$ the Friedman coordinate.
In the FLRW spacetime therefore, the Raychaudhuri equation can be rewritten as
\begin{align}
\label{Rmn2}
3(\overline{H}^2+\dot{\overline{H}})+\overline{H}\frac{\theta}{a}+\frac{\dot{\theta}}{a}+\frac{1}{3}\frac{\theta^2}{a^2}=3(H^2+\dot{H}).
\end{align}
Terms on the left hand side of Eq. (\ref{Rmn2}) are composed of background quantities, while terms on the right hand side of Eq. (\ref{Rmn1}) are composed of local perturbed quantities. Equation (\ref{Rmn2}) can be reexpressed as

\begin{align}
\label{Rmn3}
\overline{H}\frac{\theta}{a}+\frac{\dot{\theta}}{a}+\frac{1}{3}\frac{\theta^2}{a^2}=3(H^2+\dot{H})-3(\overline{H}^2+\dot{\overline{H}})=-\frac{1}{2}k^2[(\rho-\overline{\rho})+3(P-\overline{P})].
\end{align}
To obtain Eq. (\ref{Rmn3}), nothing but a FLRW spacetime filled with a perfect fluid is assumed; hence it can be applied to nonstandard cosmologies. In particular, no connection between the energy content and the geometry has been assumed, so we can study the evolution of the density perturbation by means of the continuity equation and the Friedmann equation, while not needing to use Einstein's equations.

In a matter-dominated universe, the continuity equation for a nonrelativistic fluid ($\rho\gg P$) is given by \cite{Peebles}
\begin{align}
\frac{d}{d\tau}\delta(\vec{x},\tau)+\nabla\cdot[(1+\delta(\vec{x},\tau))\vec{\upsilon}(\vec{x},\tau)]=0,
\end{align}
where $d\tau=\frac{1}{a}dt$ is the conformal time. This continuity equation can also be written as
\begin{align}
\frac{d\delta}{d\tau}+(1+\delta)\theta =0.
\end{align}
Then Eq. (\ref{Rmn3}) can be rewritten as
\begin{eqnarray}
\frac{d^2\delta}{d\tau^2}+\dot{a}\frac{d\delta}{d\tau}-\frac{4}{3}\frac{1}{1+\delta}\left(\frac{d\delta}{d\tau}\right)^2
=-3a^2(1+\delta)[(H^2+\dot{H})-(\overline{H}^2+\dot{\overline{H}})].
\end{eqnarray}
Re-scaling the time variable, $\eta=\ln a$, we obtain
\begin{eqnarray}
\label{growth}
\frac{d^2\delta}{d\eta^2}+\left(2+\frac{\dot{\overline{H}}}{\overline{H}^2}\right)\frac{d\delta}{d\eta}-\frac{4}{3}\frac{1}{1+\delta}\left(\frac{d\delta}{d\eta}\right)^2
=-3(1+\delta)\frac{(H^2+\dot{H})-(\overline{H}^2+\dot{\overline{H}})}{\overline{H}^2}.
\end{eqnarray}
Expand the $H^2+\dot{H}$ terms in term of $\delta$ and then the whole right hand side of Eq. (\ref{growth}) as
\begin{eqnarray}
\label{exp}
3(1+\delta)\frac{(H^2+\dot{H})-(\overline{H}^2+\dot{\overline{H}})}{\overline{H}^2}\equiv3(1+\delta)\sum_{n=1}c_n\delta^n.
\end{eqnarray}
To study how a small perturbation grows with time at different orders in perturbation theory, we expand $\delta$ as
\begin{eqnarray}
\delta=\sum^\infty_{i=1}\delta^i=\sum^\infty_{i=1}\frac{D_i(\eta)}{i!}\delta^i_0,
\end{eqnarray}
where $\delta_0$ is the small perturbation. According to this equation and expanding the perturbation in Eq. (\ref{exp}), we obtain the linear equation
\begin{eqnarray}
\label{linear}
D''_1+\left(2+\frac{\dot{\overline{H}}}{\overline{H}^2}\right)D'_1+3c_1D_1=0,
\end{eqnarray}
where the prime denotes a derivative with respect to the natural logarithm of the scale factor. For $\Lambda$CDM, the linear equation (\ref{linear}) reduces to
\begin{eqnarray}
D''_1+\left[2-\frac{3}{2}\frac{\Omega_{\rm m0}}{\Omega_{\rm m0}+\Omega_{\Lambda0}a^3}\right]D'_1-\frac{3}{2}\frac{\Omega_{\rm m0}}{\Omega_{\rm m0}+\Omega_{\Lambda0}a^3}D_1=0,
\end{eqnarray}
where $\Omega_{\rm m0}\equiv 8\pi G\rho_0/(3H^2_0)$ and $\Omega_{\Lambda0}\equiv \Lambda/(3H^2_0)$. At early times, $a\ll 1$, the $\Lambda$CDM universe reduces to the Einstein-de Sitter (EdS) universe ($\Omega_{\Lambda0}=0$) with solutions
\begin{eqnarray}
\label{eds}
D_1=C_1a+C_2a^{-3/2};
\end{eqnarray}
here and in the following content $C_i$ are integral constants. These solutions reproduce the usual linear growth $D_1\sim a$ and the decaying solutions $D_1\sim a^{-3/2}$.
The second order equation is found to be
\begin{eqnarray}
\label{sec}
D''_2+\left(2+\frac{\dot{\overline{H}}}{\overline{H}^2}\right)D'_2-\frac{8}{3}(D'_1)^2+3c_1D_2+6(c_1+c_2)D^2_1=0.
\end{eqnarray}
One can go on to arbitrary order by using the solutions to the lower order equations. For $\Lambda$CDM, the second order equation (\ref{sec}) reduces to
\begin{eqnarray}
D''_2+\left[2-\frac{3}{2}\frac{\Omega_{\rm m0}}{\Omega_{\rm m0}+\Omega_{\Lambda0}a^3}\right]D'_2-\frac{8}{3}(D'_1)^2-\frac{3}{2}\frac{\Omega_{\rm m0}}{\Omega_{\rm m0}+\Omega_{\Lambda0}a^3}D_2
-\frac{3\Omega_{\rm m0}}{\Omega_{\rm m0}+\Omega_{\Lambda0}a^3}D^2_1=0.
\end{eqnarray}
For Gaussian initial conditions, the skewness of the density field at large scale is related to the second-order equations, defined as
\begin{eqnarray}
S_3=3\frac{D_2}{D^2_1}.
\end{eqnarray}
For an EdS universe the skewness can be calculated analytically: $S^{\rm EdS}_3=\frac{34}{7}$.

\section{Superfluid Chaplygin gas model with FLRW spacetime}
It was argued in \cite{Popov} that SCG does not depend on details of microscopic structure of the quantum liquid and then capitalizes on effective macroscopic quantities.
SCG involves two independent flows: the coherent motion of the ground state named superfluid component, and a normal component produced by the quasiparticle gas. The particle number current and the energy-momentum tensor are, respectively, represented as
\begin{eqnarray}
&& n^{\mu}=n_{\rm c}V^{\mu}+n_{\rm n}U^{\mu},\\
&& T_{\mu\nu}=\mu~n_{\rm c}V_{\mu}V_{\nu}+W_{\rm
n}U_{\mu}U_{\nu}-Pg_{\mu\nu},
\end{eqnarray}
where $\mu$ is the chemical potential, $n_{\rm c}$ and $n_{\rm n}$
are the particle density of the superfluid and of the normal component respectively, $V^{\mu}$ and $U^{\mu}$ are the unit 4-velocities of the superfluid and the normal component.
In general, a cosmological model in which two perfect fluids flow with distinct 4-velocities should give rise to anisotropic pressures \cite{Letelier,Letelier1986,Coley},
and it has been shown that the universe would acquire some anisotropic characteristics and its geometry will deviate from the standard FLRW one if there is a slight difference between the 4-velocities of the dark energy and dark matter \cite{Harko}. However, if the two 4-velocities are parallel, which is the case we are interested here, the spacetime can be homogeneous. Second, the observational data do not forbid the anisotropic evolution at the early stage but constraints are considerable, so in any realistic model the effect of the anisotropy at the early stage may disappear rapidly. The consideration of an anisotropic behavior is useful to study all sides of the model and to get constraints to its parameters, but this topic is beyond the scope of this paper. The generalized pressure is assumed to be with the form
\begin{eqnarray}
P(\mu,\beta,\gamma)=p_{\rm c}(\mu)+p_{\rm n}(\mu,\beta,\gamma),
\end{eqnarray}
where $\gamma=V^{\mu}U_{\mu}$ associated with the relative motion of
the components, and $\beta$ is the inverse temperature with respect to the reference
frame comoving to the excitation gas. The excited state is
described by the relations
\begin{eqnarray}
\label{EoSm}
\mu n_{\rm n}=\gamma(1-c^2_{\rm s})W_{\rm
n},~~~~~p_{\rm n}=\frac{c^2_{\rm s}W_{\rm n}}{1+\nu},
\end{eqnarray}
where the adiabatic speed of sound is $c^2_{\rm s}=4M^2/\rho_{\rm
c}^2$ with $M$ a constant, and $\nu$ is a properly polytropic index. The
assumption of $p_{\rm n}\propto T^{\nu+1}$ is a generalization of
the dependence $p_{\rm n}\propto T^4$ followed by the relation
$p_{\rm n}=c^2_{\rm s}W_{\rm n}/4$. The background superfluid obeys
\begin{eqnarray}
n_{\rm c}=\sqrt{\frac{\lambda}{M}}\sqrt{\rho_{\rm
c}^2-4M^2},~~~~~p_{\rm c}=-\frac{4M^2}{\rho_{\rm c}},
\end{eqnarray}
In a spatially flat, homogeneous and isotropic FLRW universe, the
superfluid and the normal velocities are equal ($\gamma=1$). Then
Einstein equations take the forms
\begin{eqnarray}
\label{H}
&&H^2=\frac{8\pi~G}{3}\rho_{\rm tot},\\
\label{acc}
&&\frac{\ddot{a}}{a}=-\frac{4\pi~G}{3}(\rho_{\rm tot}+p_{\rm tot}),
\end{eqnarray}
where $\rho_{\rm tot}=\rho_{\rm c}+\rho_{\rm n}$ with
$\rho_{\rm n}=W_{\rm n}-p_{\rm n}$, and
$p_{\rm tot}=p_{\rm c}+p_{\rm n}$. The local energy-momentum
conservation $\nabla_\mu T^{\mu\nu}=0$ and the particle number
conservation $\nabla_\mu n^{\mu\nu}=0$, respectively, leads to
\begin{eqnarray}
\label{mc}
&&\dot{\rho}_{\rm tot}=-3H(\rho_{\rm tot}+p_{\rm tot}),\\
\label{pc} &&\dot{n}_{\rm tot}=-3Hn_{\rm tot}.
\end{eqnarray}
Taking into account Eqs. (\ref{EoSm}) and (\ref{pc}), we can rewrite
Eqs. (\ref{H}) and (\ref{mc}) as the following forms:
\begin{eqnarray}
\label{H1}
&&H^2=\frac{1}{3(1+\nu)}\left[\frac{1}{\rho}+\frac{k}{a^3}\left(\frac{\nu\rho}{\sqrt{\rho^2-1}}
+\frac{\sqrt{\rho^2-1}}{\rho}\right)\right],\\
\label{mc1}
&&3H\left(1+\nu-\frac{k}{a^3}\frac{1}{\sqrt{\rho^2-1}}\right)+\frac{\dot{\rho}}{\rho}
\left[1-\frac{k}{a^3}\left(\frac{1}{\sqrt{\rho^2-1}}-\frac{\nu\rho^2}{(\rho^2-1)^{3/2}}\right)\right]=0,
\end{eqnarray}
where $\rho=\rho_{\rm c}/(2M)$ and $k=n_0/(2\sqrt{M\lambda})$. Given
any value of the parameter $\nu$, one can have a different solution
for the fluid dynamics; see, for example, that the quasiparticle behaves
like dust ($p_{\rm n}=0$) in the limit $\nu\rightarrow \infty$, which is the case we are interested in here. In this case, Eqs. (\ref{H1}) and (\ref{mc1}) can be
solved analytically and yield to
\begin{eqnarray}
\label{mc2}
\rho_{\rm c}=\sqrt{1+\frac{k^2}{(a^3+k_0)^2}},\\
\rho_{\rm n}=\frac{k_0}{a^3}\sqrt{1+\frac{k^2}{(a^3+k_0)^2}}.
\end{eqnarray}
With these two equations, the Eq. (\ref{H1}) turns out to be
\begin{eqnarray}
\label{H3}
H^2=\frac{8\pi G}{3}\left(1+\frac{k_0}{a^3}\right)\sqrt{1+\frac{k^2}{(a^3+k_0)^2}}.
\end{eqnarray}
Equation (\ref{acc}) can be rewritten as
\begin{eqnarray}
\label{H4}
\dot{H}=-4\pi G\left[1+\frac{k_0}{a^3}-\frac{(a^3+k_0)^2}{k^2+(a^3+k_0)^2}\right]\sqrt{1+\frac{k^2}{(a^3+k_0)^2}}.
\end{eqnarray}
If we define $E^2=H^2/H^2_0$ and $\Omega_{\rm c}=8\pi G\rho_{\rm
c0}/(3H^2_0)$, since $E(z=0)=1$, there is a relation between $\Omega_{\rm
c}$ and $k_0$: $\Omega_{\rm c}=1/(1+k_0)$.

\section{Gravitational growth and large structure in SCG scenario}
In the following, we will discuss the gravitational growth and constraints coming from the linear and nonlinear aspects of structure formation in the SCG scenario, such as $S_3$ the skewness of the density field and $\sigma_8$ the rms fluctuations on a sphere of 8 Mpc $h^{-1}$ by using the analysis presented above.
\subsection{Gravitational growth}
Assume that the dark energy component of SCG affects large-scale structure growth only through its effect on the background evolution, while the fluctuations of the dark matter component of SCG are responsible for the large-scale structure. According to Eq. (\ref{exp}), we get
\begin{eqnarray}
&& c_1=-\frac{4\pi G}{3}\frac{\overline{\rho}_n}{\overline{H}^2}=-\frac{1}{2}\frac{k_0}{a^3+k_0},\\
&& c_2=0.
\end{eqnarray}
From Eqs. (\ref{H3}) and (\ref{H4}), the term appearing in front of the $\delta'$ in the perturbation Eq. (\ref{growth}) is found to be
\begin{eqnarray}
2+\frac{\dot{\overline{H}}}{\overline{H}^2}=\frac{1}{2}+\frac{3}{2}\frac{a^3(a^3+k_0)}{k^2+(a^3+k_0)^2}.
\end{eqnarray}
At early times ($a\ll 1$), SCG reduces to the EdS (also $\Lambda$CDM) case with solution given by equation (\ref{eds}). In the large $a$ limit, the linear equation (\ref{linear}) in SCG reduces to \begin{eqnarray}
D''_1+2D'_1-\frac{3}{2}\frac{k_0}{a^3}D_1=0.
\end{eqnarray}
Accordingly, the solution to this equation is
\begin{eqnarray}
\label{sol1}
D_1=\frac{c_3}{a}F_1\left(\frac{2}{3},\xi(a)\right)
+\frac{c_4}{a}F_2\left(\frac{2}{3},\xi(a)\right),
\end{eqnarray}
where $F_1$ and $F_2$ are the Bessel functions of the first and second kinds with $\xi(a)=\frac{1}{3}i\sqrt{6k_0}a^{-3/2}$, respectively.

\begin{figure}
\includegraphics[width=10cm]{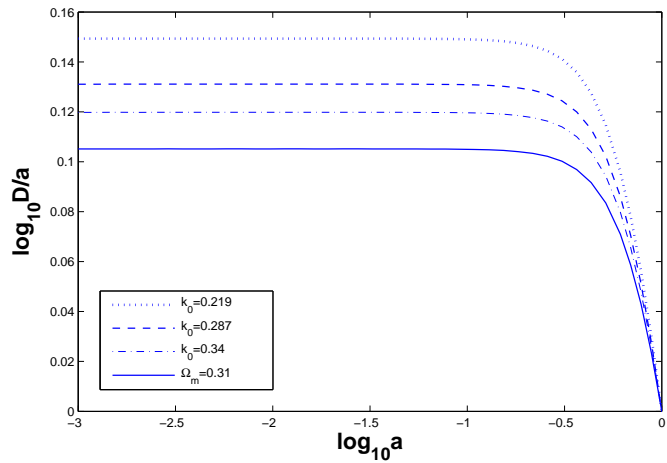}
\caption{Linear growth $D$ (normalized with the scale factor $a$) for $\Omega_{\rm m0}=0.31$ \cite{Planck} in $\Lambda$CDM and different values of $k_0$ with $k=0.173$ in SCG. \label{growth1}}
\end{figure}

\begin{figure}
\includegraphics[width=10cm]{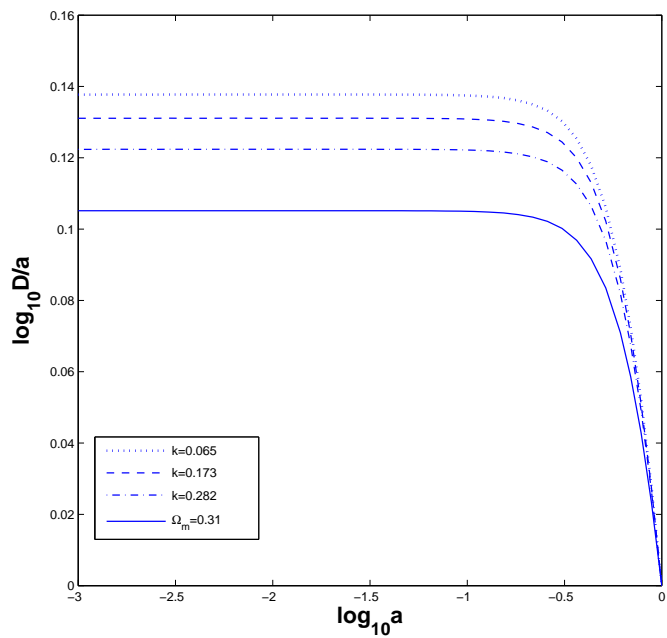}
\caption{Linear growth $D$ for $\Omega_{\rm m0}=0.31$ in $\Lambda$CDM and different values of $k$ with $k_0=0.287$ in SCG. \label{growth2}}
\end{figure}

In Figs. \ref{growth1} and \ref{growth2}, we plot numerically the linear growth factor $D=D_1(z)/D_1(0)$ of the perturbations for different values of $k_0$ (with $k=0.173$ obtained in \cite{Lazkoz}) and $k$ (with $k_0=0.287$ obtained in \cite{Lazkoz}). The linear growth is normalized with the scale factor, $a$, the growth rate in the EdS universe (it is also equal to the growth rate in the $\Lambda$CDM universe at early times). The differential equations are solved numerically in the region of cosmological interest $a=0.001$ to $a=1$. The initial condition is chosen such that at $a=0.001$ the standard exponential solution, $D_1\sim a$, is reached. The form of the linear growth factor is similar to the case of $\Lambda$CDM. The smaller the value of $k_0$ (or $k$) is, the larger the deviation of the growth factor between SCG and $\Lambda$CDM is, but the variation is observationally insignificant, being at best less than 4\%. There will be more stricture growth on the large scale in the SCG scenario than in $\Lambda$CDM.

\subsection{Amplitude normalization: $\sigma_8$}
Another important parameter related to the growth is the $\sigma_8$ parameter which is the rms matter density contrast in a sphere with a comoving radius of $8h^{-1}$ Mpc at present, where $h$ is the usual dimensionless Hubble constant in units of $100$ km~s$^{-1}$~Mpc$^{-1}$. The rms mass fluctuation $\sigma_8$ is defined as
\begin{eqnarray}
\label{sol1}
\sigma^2(R,z)=\int_0^{\infty}\frac{{\rm d}k}{k}W^2(kR)\Delta^2(k,z),
\end{eqnarray}
where $W(kR)$ is the window function defined as

\begin{eqnarray}
W(kR)=3\left[\frac{\sin(kR)}{(kR)^3}-\frac{\cos(kR)}{(kR)^2}\right]r,
\end{eqnarray}
and
\begin{eqnarray}
\Delta^2(k,z)=4\pi k^3P_\delta(k,z),
\end{eqnarray}
where $P_\delta(k,z)$ is the primordial matter power spectrum. The function $\sigma_8$ is related to $D(z)$ as
\begin{eqnarray}
\sigma_{8}(z)=D(z)\sigma_{8}(0).
\end{eqnarray}
Here the value of $\sigma_{8}(0)$ is normalized to the $\Lambda$CDM model according to $\sigma_{8}(0)=\frac{D_{1}(0)}{D_{1,\Lambda\rm{CDM}}(0)}\sigma_{8, \Lambda\rm{CDM}}(0)$ with $\sigma_{8, \Lambda\rm{CDM}}(0)=0.83$ \cite{Planck}.

\begin{figure}
\includegraphics[width=10cm]{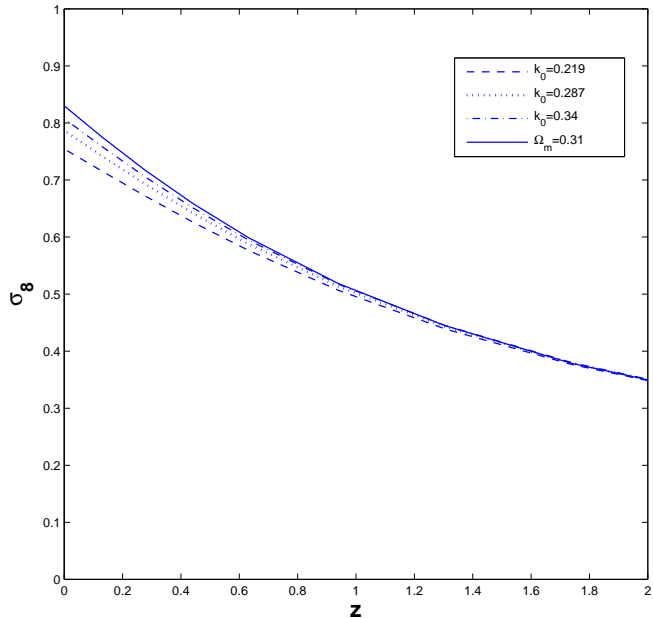}
\caption{The evolution of $\sigma_8(z)$ with different values of $k_0$ and $k=0.173$ in SCG, comparing with the $\Lambda$CDM case. \label{sigma81}}
\end{figure}

\begin{figure}
\includegraphics[width=10cm]{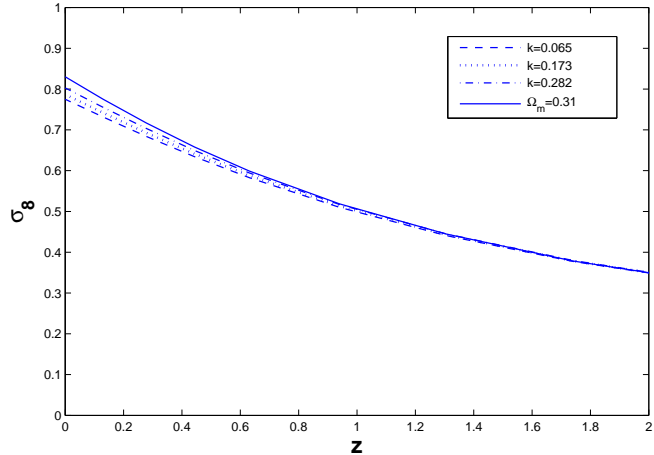}
\caption{The evolution of $\sigma_8(z)$ with different values of $k$ and $k_0=0.287$ in SCG, comparing with the $\Lambda$CDM case. \label{sigma82}}
\end{figure}

In Figs. \ref{sigma81} and \ref{sigma82}, the evolution of $\sigma_{8}(z)$ in the SCG scenario is plotted numerically. We can see that the value of $\sigma_{8}(0)$ is very close to $0.8$ when the value of $k_0$ (or $k$) increases. We also plot $\sigma_{8, \Lambda\rm{CDM}}(z)$ for comparison. The variation between $\sigma_{8}(z)$ and $\sigma_{8, \Lambda\rm{CDM}}(z)$ almost cannot be discriminated at $z\sim 1$. The deviation is observationally insignificant even at $z=0$, being at best less than 6\% when $k_0$ (or $k$) takes values changing from the best-fit value in the $1~\sigma$ range obtained in \cite{Lazkoz}. In other words, the value of $\sigma_{8}(0)$ is compatible with that of $\Lambda$CDM, and we cannot use $\sigma_{8}(z)$ to distinguish the SCG model from the $\Lambda$CDM model at $68.3\%$ confidence level. We can conclude, however, that $\sigma_{8}(0)$ in the SCG scenario is smaller than that of the $\Lambda$CDM model at $68.3\%$ confidence level. Significantly smaller $\sigma_{8}$ have been found at low redshift by velocity fields \cite{Willick} and some weak lensing studies \cite{Jarvis}, which can be more easily accounted for by using the SCG cosmological model.

\subsection{The skewness $S_3$}
Now we investigate the nolinear aspects of the large-scale structure formation, such as the skewness of the density field. In the large $a$ limit, the second order equation (\ref{sec}) in SCG reads
\begin{eqnarray}
\label{2scg}
D''_2+2D'_2-\frac{8}{3}(D'_1)^2-\frac{3k_0}{2a^3}D_2-\frac{3k_0}{a^3}D^2_1=0.
\end{eqnarray}
Substituting the Eq. (\ref{sol1}), the solution to this equation is
\begin{eqnarray}
D_2&=&\frac{c_5}{a}F_1\left(\frac{2}{3},\xi(a)\right)+\frac{c_6}{a}F_2\left(\frac{2}{3},\xi(a)\right)+\frac{\pi k_0}{3a}\Bigg[F_1\left(\frac{2}{3},\xi(a)\right)\int\bigg(-\frac{4}{a^4}F_2\left(\frac{2}{3},\xi(a)\right) \bigg( c^2_3 F^2_1\left(-\frac{1}{3},\xi(a)\right)\\\nonumber
&&-\frac{3c^2_3}{4}F^2_1\left(\frac{2}{3},\xi(a)\right)+2c_3c_4F_1\left(-\frac{1}{3},\xi(a)\right)F_2\left(-\frac{1}{3},\xi(a)\right)-\frac{3}{2}c_3c_4F_1\left(\frac{2}{3},\xi(a)\right)F_2\left(\frac{2}{3},\xi(a)\right)\\\nonumber
&&+c^2_4F^2_2\left(-\frac{1}{3},\xi(a)\right)-\frac{3}{4}c^2_4F^2_2\left(\frac{2}{3},\xi(a)\right)\bigg)\bigg)dx -F_2\left(\frac{2}{3},\xi(a)\right)\int \Bigg(-\frac{4}{a^4}F_1\left(\frac{2}{3},\xi(a)\right)\bigg(-\frac{3c^2_3}{4}F^2_1\left(\frac{2}{3},\xi(a)\right)\\\nonumber
&&-\frac{3}{2}c_3c_4F_1\left(\frac{2}{3},\xi(a)\right)F_2\left(\frac{2}{3},\xi(a)\right)+c^2_3F^2_1\left(-\frac{1}{3},\xi(a)\right)+2c_3c_4F_1\left(-\frac{1}{3},\xi(a)\right)F_2\left(-\frac{1}{3},\xi(a)\right)\\\nonumber
&& +c^2_4\left(F^2_2\left(-\frac{1}{3},\xi(a)\right)-\frac{3}{4}F^2_2\left(\frac{2}{3},\xi(a)\right)\right)\bigg)\Bigg) dx\Bigg],
\end{eqnarray}
From the first and the second order equations, we see that a SCG universe behaves fundamentally differently from the $\Lambda$CDM and GCG universe: at early times ($a\ll 1$), SCG and $\Lambda$CDM reduce to the EdS case, whereas GCG does not; at very late times ($a\gg 1$), GCG reduces to the $\Lambda$CDM case, whereas SCG does not.

\begin{figure}
\includegraphics[width=10cm]{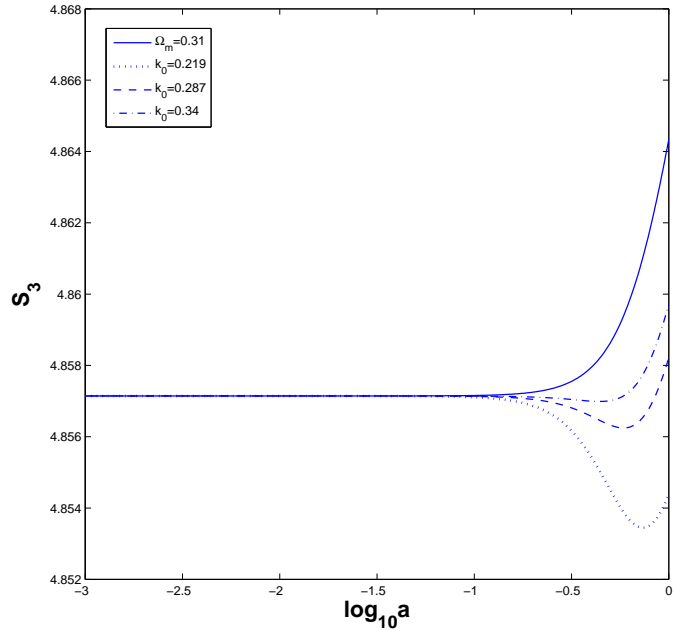}
\caption{The evolution of $S_3(z)$ with different values of $k_0$ and $k=0.173$ in SCG, comparing with the $\Lambda$CDM case.. \label{skewness1}}
\end{figure}

\begin{figure}
\includegraphics[width=10cm]{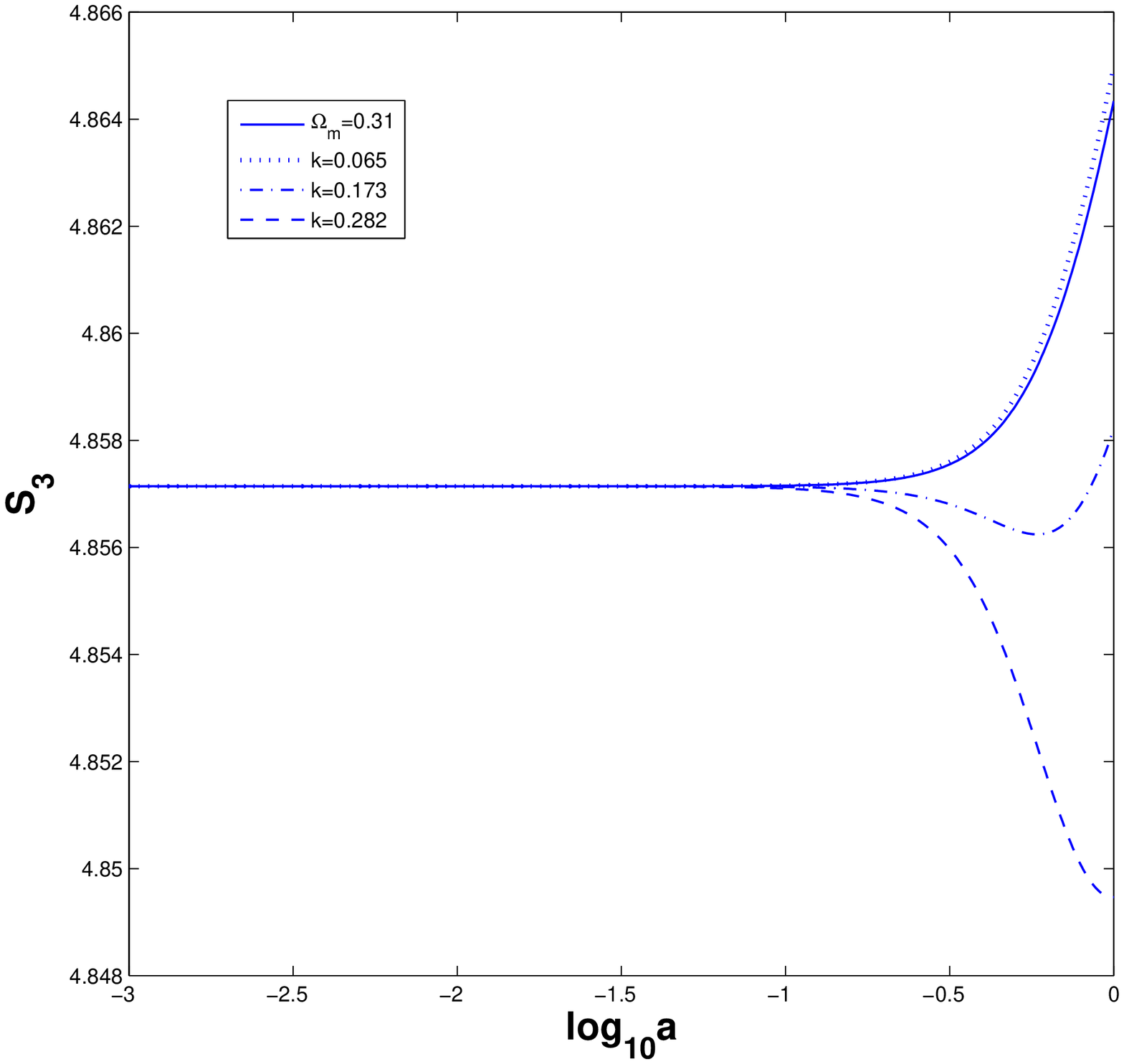}
\caption{The evolution of $S_3(z)$ with different values of $k$ and $k_0=0.287$ in SCG, comparing with the $\Lambda$CDM case. \label{skewness2}}
\end{figure}

In Figs. \ref{skewness1} and \ref{skewness2}, we plot the evolution of the skewness $S_3$ in the SCG scenario. To solve the second order equation (\ref{2scg}) numerically, we chose the initial condition as $S_3=34/7$ at $a=0.001$; namely the standard solution, the EdS universe, is recovered at the early time. It is obvious that the deviation of skewness $S_3(0)$ between the SCG and $\Lambda$CDM models is very small, being at best less than 1\% when $k_0$ (or $k$) takes values changing from the best-fit value in the $1~\sigma$ range obtained in \cite{Lazkoz}. For $a\leq 0.1$, the evolution of $S_3$ in the SCG and $\Lambda$CDM models is almost the same. Current estimations for $S_3$ agree with the standard predictions but with large uncertainties, of order $20\%-30\%$ \cite{Bernardeau2002}. Thus current observations on $S_3$ cannot be used to distinguish different cosmological models. Future measurement on $S_3$, after removing biasing uncertainties, could be used to constrain the parameters of cosmological models \cite{Reis}.
%

\section{Summary and conclusions}
The growth of the large-scale structure in SCG cosmology has been studied. SCG with two free parameters is a phenomenological viable model in which the first and the second order perturbation equations have been obtained. The linear growth factor, $\sigma_{8}$, and $S_3$ are important quantities in studies of the large-scale structure, and all of them were investigated here.

The form of the linear growth factor is similar to the $\Lambda$CDM case. There is more stricture growth on large scales in the SCG model than in $\Lambda$CDM. The smaller the values of parameters are, the larger the deviation of the growth factor between SCG and $\Lambda$CDM is, but the variation is at best less than 4\%.

The variation between $\sigma_{8}(z)$ and $\sigma_{8, \Lambda\rm{CDM}}(z)$ cannot be discriminated, being at best less than 6\% when parameters of SCG take values changing from the best-fit value in the $1~\sigma$ range obtained in \cite{Lazkoz}. The $\sigma_{8}(0)$ in the SCG scenario is smaller than that of the $\Lambda$CDM model, which can more easily account for the smaller $\sigma_{8}$ found at low redshift by some weak lensing studies and velocity fields.

We also have found that the deviation of skewness $S_3$ between the SCG and $\Lambda$CDM models is very small, being at best less than 1\% when the parameters change from the best-fit value in the $1~\sigma$ range obtained in \cite{Lazkoz}. Thus current observations of $S_3$ cannot be used to distinguish the SCG model from $\Lambda$CDM.

\begin{acknowledgments}
I thank Y. Liu for helpful discussions. I am grateful to the Center for High Energy Physics of Peking University and the Kavli Institute for Theoretical Physics China for hospitality
where this work was completed during a program on Cosmology after Planck. This study is supported in part by National Natural Science Foundation of China (Grant Nos. 11147028 and 11273010) and Hebei Provincial Natural Science Foundation of China (Grant No. A2011201147 and A2014201068).
\end{acknowledgments}

\bibliography{apssamp}

\end{document}